\documentclass[12pt]{article}

\usepackage{amssymb}
\usepackage{color}
\usepackage{graphicx,psfrag}
\usepackage{fancybox}

\begin{document}

\centerline{\Large\bf Inversion of Gamow's Formula and Inverse
Scattering}

\vspace{5mm}
\begin{center}
   {\bfseries S.C. Gandhi and C.J. Efthimiou}\\
   Department of Physics\\
   University of Central Florida\\
   Orlando, FL 32826
\end{center}

\begin{abstract}
We present a pedagogical description of the inversion of Gamow's
tunnelling formula and we compare it with the corresponding
classical problem. We also discuss the issue of uniqueness in the
solution and the result is compared with that obtained by the
method of Gel'fand and Levitan. We hope that the article will be a
valuable source to students who have studied classical mechanics
and have some familiarity with quantum mechanics.
\end{abstract}

\section{Introduction}

Eugen Merzbacher has commented that \cite{Merzbacher} ``among all
the successes of quantum mechanics as it evolved in the third
decade of the 20th century, none was more impressive than the
understanding of the tunnel effect---the  penetration of matter
waves and the transmission of particles through a high potential
barrier." The tunnel effect provided a  straightforward and
remarkable explanation of the radioactive $\alpha$-decay of
nuclei. George Gamow was one of the---although not the sole--
protagonists in the discovery of the theory of $\alpha$-decay
\cite{Merzbacher,alphadecay} and the basic formula, equation
(\ref{eq:Gamow}), that underlies tunnelling through a potential
barrier is often referred to as---perhaps
unjustly---\textbf{Gamow's penetrability factor} (see, for
example, p. 62 of \cite{Constantinescu}).

Leaving aside Gamow's mischievous account of history, this article
will discuss how knowledge of the tunnelling behavior of a
potential can be used to determine the potential itself. The
quantum mechanical problem will also be contrasted with the
classical version in order to gain further insight.

%%%%%%%%%%%%%%%%%%%%%%%%%%%%%%%%%%%%%%%%%%%%%%%%%%%%%%%%%%%%%%%%%%%%%%%%%%%
\section{Classical vs Quantum Problem}

%%%%%%%%%%%%
\subsection{The Classical Problem}
\begin{figure}[h!]
\begin{center}
    \psfrag{E}{$E$}
    \psfrag{u}{$U$}
    \psfrag{x}{$x$}
    \psfrag{m}{\textcolor{blue}{$m$}}
    \psfrag{x1(u)}{\textcolor{red}{$x=x_1(U)$}}
    \psfrag{x2(u)}{\textcolor{red}{\hspace{-5mm}$x=x_2(U)$}}
    \psfrag{x1}{$x_1$}
    \psfrag{x2}{$x_2$}
  \includegraphics[height=4.5cm]{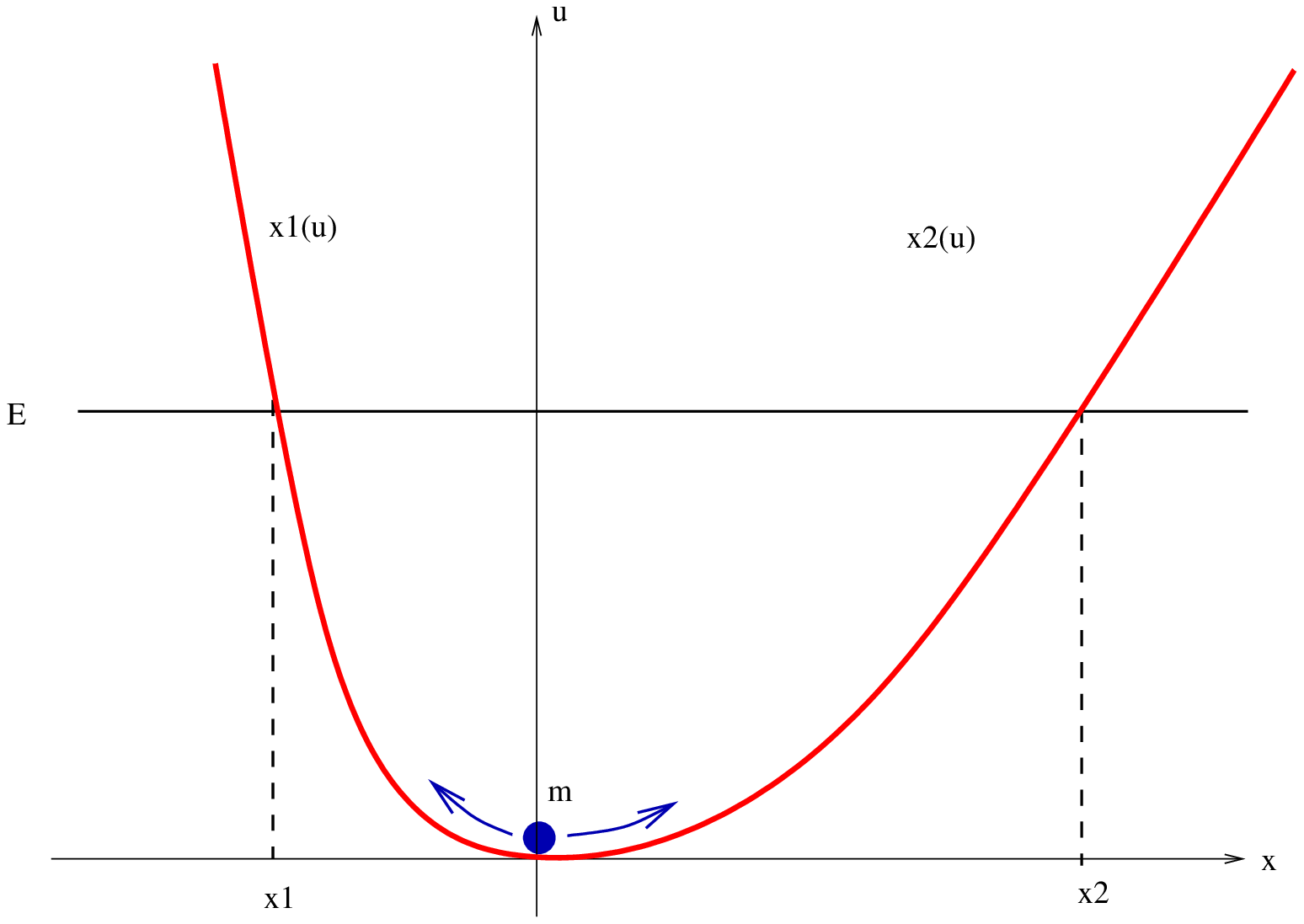}
\hspace{1cm}
    \psfrag{v}{\textcolor{blue}{$m$}}
    \psfrag{x}{$x$}
    \psfrag{u}{$U$}
    \psfrag{u0}{$U_0$}
    \psfrag{E}{$E$}
    \psfrag{0}{$0$}
    \psfrag{L}{$L$}
    \psfrag{a}{\hspace{-3mm}$x_1(E)$}
    \psfrag{b}{\hspace{-2mm}$x_2(E)$}
    \includegraphics[height=4.5cm]{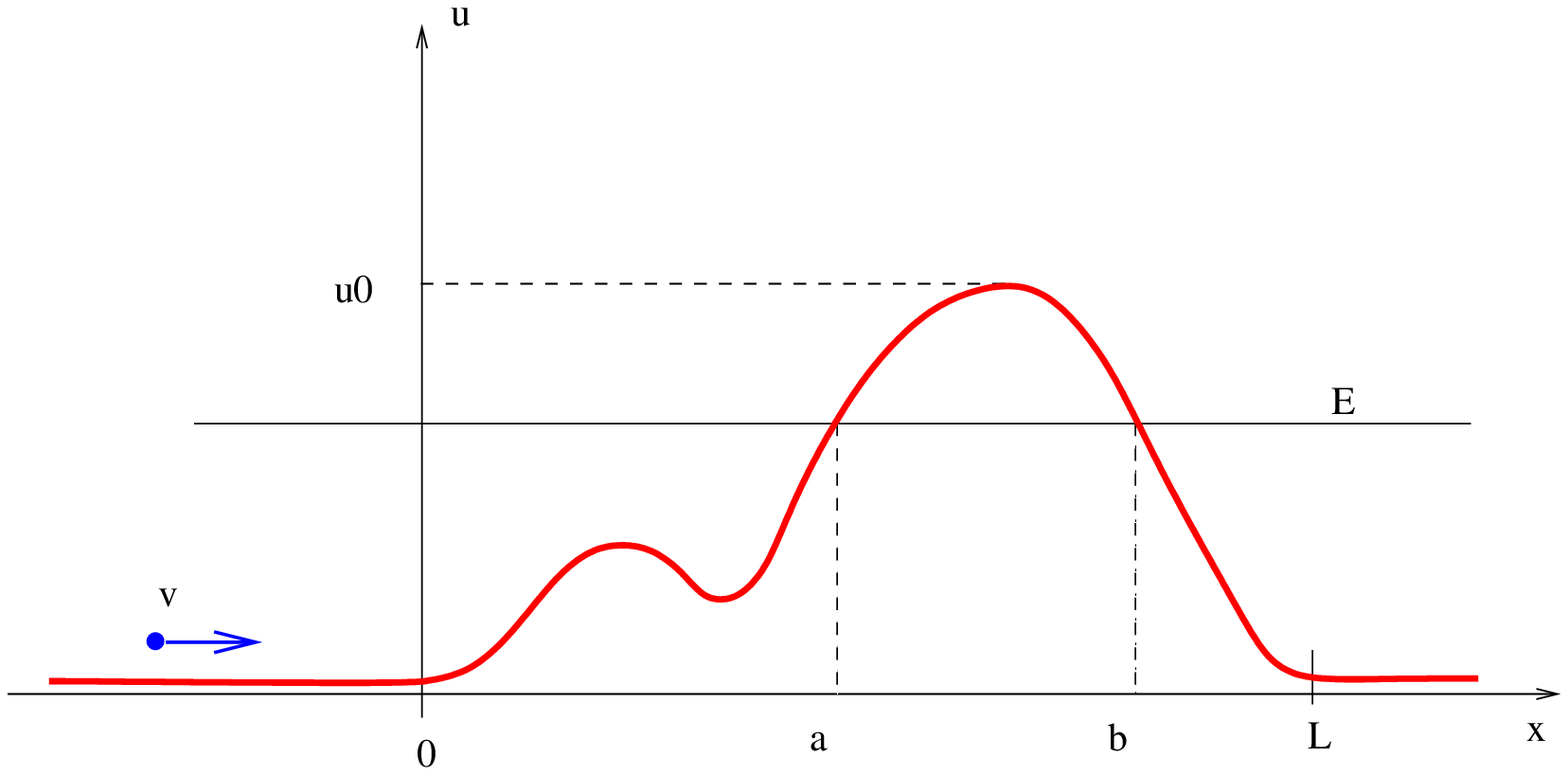}
  \end{center}
  \caption{A particle in a potential well (attractive force field) performs oscillations;
           a particle on a potential barrier (repulsive force field) will either overcome
           the potential if it has enough energy ($E>U_0$) or will reflect back if it
           does not have enough energy $E<U_0$.}
  \label{classical}
\end{figure}

The classical, one-dimensional inverse scattering problem is
probably best exemplified by the two systems depicted in figure
\ref{classical}.

 To the left of figure \ref{classical} is a particle oscillating in an
attractive potential with single minimum, set to zero and
occurring at the origin. Conservation of energy gives the period,
\begin{displaymath}
  T(E) ~=~ \sqrt{2m}\int_{x_1(E)}^{x_2(E)}{\mathrm{d}x \over
  \sqrt{E-U(x)}}~,
\end{displaymath}
where $x_1(E)$ and $x_2(E)$ are the turning points for the energy
$E$. The inverse problem is to determine the form of the potential,
$U(x)$, given the period as a function of energy, $T(E)$.  The
solution is well known and may be found in many standard texts on
classical mechanics (see, for example \cite{landau}).  Treating $x$
as a function of $U$ rather than $U$ a function of x, the above may
be turned into Abel's integral equation.  Since $U(x)$ is not
one-to-one this requires splitting its domain at the origin and
defining the two functions $x_1(U)$ and $x_2(U)$ as per figure
\ref{classical}. The result is
\begin{displaymath}
  x_2(U)-x_1(U) ~=~ {1 \over \pi\sqrt{2 m }}\int_0^U{{T(E) \mathrm{d} E
 \over \sqrt{U-E}}}~.
\end{displaymath}
We find that the solution cannot be determined uniquely unless the
additional, assumption that the potential is even, is introduced.
%about but only to within a symmetry family.
In this case
\begin{displaymath}
  x(\tilde U) ~=~ {1 \over 2\pi\sqrt{2 m }}\int_0^{\tilde U}{{T(E) \mathrm{d} E
 \over \sqrt{\tilde U-E}}}~,
\end{displaymath}
where we have denoted the unique, even solution by $\tilde U(x)$.

To the right of figure \ref{classical} is a particle incident on a
potential barrier that is confined to the interval $[0,\,L]$.  The
problem of determining the potential given the time of traversal of
the potential as a function of energy has been solved by Lazenby and
Griffiths \cite{griffiths}.  The forward and backward scattering
data are defined respectively as,
\begin{eqnarray*}
 T(E) &=& \sqrt{{m \over 2}}\int_0^L{{\mathrm{d} x \over \sqrt{E-U(x)}}}~,
           ~~~~~~~\mbox{if}~E > U_0~,\\
 R(E) &=& \sqrt{{m \over 2}}\int_0^{x_1(E)}{{\mathrm{d} x\over\sqrt{E-U(x)}}}~,
           ~~~\mbox{if}~E \leq U_0~,
\end{eqnarray*}
where $x_1(E)$ is the left turning point. The former applies when
the particle has energy exceeding $U_0$, the maximum of the
potential, and gives the time required for the particle to
traverse the potential.  The latter applies when $E\leq U_0$ and
gives the time required for the particle to reach the turning
point $x_1(E)$ or half the time taken for the particle to return
to the origin.

The solutions to the barrier equations are quite similar to that
of the previous system. The most important feature is that a class
of potentials are obtained. There is, however, a unique solution
with the property that it increases monotonically over the
interval $[0,L]$ (and drops discontinuously to zero at $L$).
Lazenby and Griffiths call this the \textit{canonical} potential
and use it to represent the class of solutions. For instance the
inversion of the backward scattering data is given by
\begin{displaymath}
   x(\tilde{U}) ~=~ {1 \over \pi} \sqrt{{2 \over m}}
  \int_0^{\tilde{U}}{{R(E) \mathrm{d} E \over \sqrt{\tilde{U} - E}}}~,
\end{displaymath}
where $\tilde{U}$ is the canonical potential.

In their paper, Lazenby and Griffiths remark that it is curious that
the above solutions are not determined uniquely whereas in the
quantum mechanical analogue the solution is unique.  It is further
remarked that ``given the transmission coefficient $T$ (the
probability that the particle will surpass the barrier) as a
function of energy $E$, (the potential) may be recovered by the
method of Gel'fand and Levitan". As will be discussed, this
statement is not accurate as it stands. The transmission coefficient
alone is not sufficient to determine the potential. The transmission
amplitude however is a complex function and carries more
information. All this will be clarified in the following.

%%%%%%%%%%%%
\subsection{The Quantum Mechanical Problem}
\begin{figure}[h!]
\begin{center}
    \psfrag{x}{$x$}
    \psfrag{u}{$U$}
    \psfrag{i}{$e^{i k x}$}
    \psfrag{r}{$e^{-i k x}$}
    \psfrag{t}{$e^{i k x}$}
    \psfrag{B}{Bound States, $E < 0$}
    \includegraphics[width=.5 \textwidth]{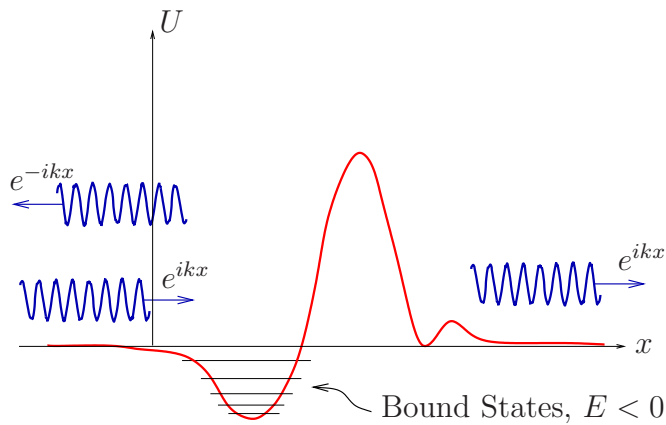}
 \caption{Quantum mechanical scattering of a right-moving particle that approaches a potential
          from left. It is assumed that the potential vanishes at large distances and,
          therefore, bound states appear only if there is a region with $U<0$ and
          appear only for negative energies.}
 \label{qm}
\end{center}
\end{figure}

The quantum mechanical problem is depicted in figure \ref{qm}.

A particle is incident from the left, say, on a potential which
limits to a constant as $x\to\pm\infty$, which we will set to
zero.  If the potential approaches zero (a constant) rapidly
enough, the asymptotic form of the wavefunction for
$x\to\pm\infty$ are plane waves:
\begin{eqnarray*}
 \psi(x) \sim \left\{ \begin{array}{ll}
    e^{i k x} + b(k)\, e^{-i k x}~, &\; \;x \to -\infty~,\\
    a(k)\, e^{i k x}~,&\;\; x \to +\infty~,
    \end{array} \right.
\end{eqnarray*}
where the energy of the particle $E>0$ and
 $$
    k ~=~ \sqrt{{2 m E \over\hbar^2}}~.
 $$
If there exits an interval over which $U(x) < 0$, then there is a
discrete spectrum $E_n$ corresponding to bound states:
\begin{eqnarray*}
 \psi(x) \sim \left\{ \begin{array}{ll}
    c_n \, e^{+\kappa_n x}~, &\; \;x \to -\infty~,\\
    d_n \, e^{-\kappa_n x}~,&\;\; x \to +\infty~,
    \end{array} \right.
\end{eqnarray*}
where $E_n<0$ and
 $$
    \kappa_n ~=~ \sqrt{{- 2 m E_n \over \hbar^2}}~.
 $$

The scattering data for the inverse problem is comprised of the
asymptotic coefficients $b(k)$ and $c_n$ as well as the discrete
eigenvalues $\kappa_n$.  The potential is then uniquely
constructed by way of the method of Gel'fand and Levitan
\cite{gelf} outlined in figure \ref{fig:GelfandLevitan} below.
\begin{figure}[h!]
 $$
  \fbox{
     \begin{Beqnarray*}
    & b(k),\;(c_n,\, \kappa_n) & \\
     &\downarrow& \\
    &F(X)=\sum{c_n^2 e^{-\kappa_n X}} + {1 \over 2 \pi}
    \int_{-\infty}^{+\infty}{b(k) e^{i k X} \mathrm{d}k}& \\
     &\downarrow& \\
    & K(x,z)+F(x+z)+\int_x^{+\infty}{K(x,y)F(y+z)\mathrm{d}y}=0& \\
     &\downarrow& \\
    &U(x) = -2\, {\mathrm{d} \over \mathrm{d} x} K(x,x)&
     \end{Beqnarray*}
         }
$$
 \caption{The Gel'fand and Levitan method in a nutshell.}
 \label{fig:GelfandLevitan}
\end{figure}

The scattering data is used to construct the auxiliary function
$F(X)$.  The auxiliary function is then used to define a linear
Fredholm integral equation, called the \textbf{Marchenko equation},
for a second auxiliary function $K(x,z)$ (third entry). The
potential is then determined by taking the directional derivative
along the line $z=x$.

%%%%%%%%%%%%%%%%%%%%%%%%%%%%%%%%%%%%%%%%%%%%%%%%%%%%%%%%%%%%%%%%%%%%%%
\section{Gamow's Formula and its Inversion}

\subsection{Gamow's Formula}

\begin{figure}[h!]
 \begin{center}
     \psfrag{u0}{$U_0$}
    \psfrag{u}{$U$}
    \psfrag{x1}{\hspace{-3mm}$x_1(E)$}
    \psfrag{x2}{\hspace{-2mm}$x_2(E)$}
    \psfrag{E}{$E$}
    \psfrag{x}{$x$}
    \includegraphics[height=4cm]{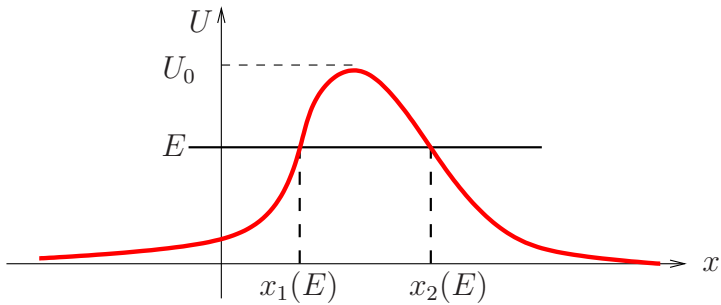}
  \end{center}
  \caption{A particle incident on a potential barrier has a finite probability to overcome
           the barrier even if its energy $E$ is below the maximum $U_0$ of the barrier.}
  \label{fig:QM}
\end{figure}

The applicable situation is illustrated in figure \ref{fig:QM}.  A
particle is incident on a potential barrier with a \textit{single}
maximum, $U_0$ (at $x=0$, say), with energy $E$ less than $U_0$.
It is known that in quantum mechanics there can be a finite
probability for the particle to exceed the potential despite the
fact that $E<U_0$.

Gamow's tunnelling formula gives a good approximation to the
transmission coefficient, $T(E)$, giving the probability for the
particle to surpass the barrier.
 \begin{equation}
   T(E) ~=~ \exp\left({-{2 \over \hbar}
            \int_{x_1(E)}^{x_2(E)}{\sqrt{2m(U(x)-E)}\,\mathrm{d}x}}\right)~.
 \label{eq:Gamow}
 \end{equation}
This formula can be proved easily by considering the barrier as an
infinite sum of infinitely thin rectangular barriers (p. 219 of
\cite{Zettili}). However, this method, although it provides the
correct result, is mathematically inconsistent. A mathematically
sound proof can be given using the JWKB approximation (p. 507 of
\cite{Zettili}).

In what follows, $T(E)$ will play the role analogous to the
classical scattering data.

%%%%%%%%%%%%%%%%%%%%%%%%%%%%%%%%%
\subsection{Inversion of Gamow's Formula}

\begin{figure}[h!]
\begin{center}
    \psfrag{a}{$\alpha$}
    \psfrag{b}{$U_0$}
    \psfrag{E}{$E$}
    \psfrag{U}{$U$}
    \psfrag{E=U}{$E=U$}
    \psfrag{II}{$\mathcal{D}$}
    \psfrag{I}{}
    \includegraphics[width=6cm]{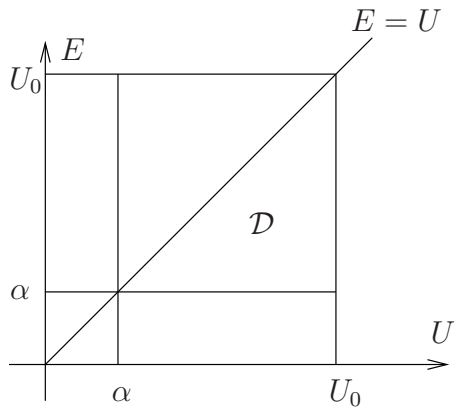}
 \caption{Our integral is defined on the triangular domain $\mathcal{D}$.}
 \label{fig:integration}
\end{center}
\end{figure}

Let us take up the task of inverting Gamow's formula.  By
differentiating and, once again, rewriting in terms of the inverse
functions $x_1(U)$ and $x_2(U)$ (we have again split the domain of
$U(x)$ at the origin), we find
\begin{equation}
 {\hbar \over \sqrt{2 m}}{1 \over T(E)}{\mathrm{d}T \over\mathrm{d}E}
  ~=~ \int_{x_1(E)}^{x_2(E)}{{\mathrm{d}x \over \sqrt{U-E}}} =
 \int_E^{U_0}{\left({\mathrm{d}x_1 \over \mathrm{d}U} -
 {\mathrm{d}x_2 \over \mathrm{d}U}\right){\mathrm{d}U \over
 \sqrt{U-E}}}~.
 \label{eq:newAbel}
\end{equation}
The above equation is nearly in the form of Abel's equation
$$
  \int_0^E {\phi(U)\over\sqrt{E-U}}\,dU ~=~ f(E)~.
$$
However, it differs in that the position of the parameter and the
variable have been switched in the root, and in that the limits of
the integral are from the variable to a constant rather than zero to
the variable:
$$
  \int_E^a {\phi(U)\over\sqrt{U-E}}\,dU ~=~ f(E)~.
$$
 Consequently, the preferred approach
of applying the Laplace transform to the equation and making use
of the convolution theorem fails.  However, Abel's equation can be
solved by composition with a kernel \cite{landau}. We can, in
fact, apply this process with some modification. We divide both
sides by $\sqrt{E-\alpha}$, where $0\leq \alpha \leq U_0$, and
integrate with respect to $E$ from $\alpha$ to $U_0$:
\begin{eqnarray*}
   {\hbar \over \sqrt{2 m}}\int_\alpha^{U_0}{{\mathrm{d}T/\mathrm{d}E
   \over T(E)\sqrt{E-\alpha}}\mathrm{d}E}
&=& \int_\alpha^{U_0}{{\mathrm{d}E \over
    \sqrt{E-\alpha}}\int_E^{U_0}{\left({\mathrm{d}x_1 \over \mathrm{d}U}
   - {\mathrm{d}x_2 \over \mathrm{d}U}\right){\mathrm{d}U \over
   \sqrt{U-E}}}} \\
   %%%%%%%%%%%%%%%%%%%%%%%%
&=& \int_\alpha^{U_0}{\left({\mathrm{d}x_1 \over \mathrm{d}U} -
   {\mathrm{d}x_2 \over
    \mathrm{d}U}\right)\mathrm{d}U\int_\alpha^{U}{{\mathrm{d}E \over
   \sqrt{(U-E)(E-\alpha)}}}}~,
\end{eqnarray*}
where in the second line we have changed the order of integration
(see figure \ref{fig:integration}).

In the appendix we show that
$$
  \int_\alpha^U {\mathrm{d}E \over \sqrt{(U-E)(E-\alpha)}} ~=~
  \pi~.
$$
Therefore
\begin{eqnarray*}
   {\hbar \over \sqrt{2 m}}\int_\alpha^{U_0}{{\mathrm{d}T/\mathrm{d}E
   \over T(E)\sqrt{E-\alpha}}\mathrm{d}E}
&=& \pi\int_\alpha^{U_0}{\left({\mathrm{d}x_1 \over \mathrm{d}U} -
    {\mathrm{d}x_2 \over \mathrm{d}U}\right)\mathrm{d}U}~,
\end{eqnarray*}
and finally:
\begin{eqnarray}
  x_1(U)-x_2(U)&=& - {\hbar \over
  \pi\sqrt{2m}}\int_U^{U_0}{{\mathrm{d}T(E)/\mathrm{d}E \over
  T(E)\sqrt{E-U}}\,\mathrm{d}E}~,
\label{eq:inversion}
\end{eqnarray}
where we have used the fact that $x_1(U_0)-x_2(U_0) = 0$. We see
that the solution is quite similar in form to the classical
result, and again the solution is not unique.  Rather, we have
obtained a family of potentials which all result in the same
transmission coefficient.

%%%%%%%%%%%%%%%%%%%%%%%%
\subsection{An Example: Cold Emission}
Although the photoelectric effect has become the most popular
example of emission of electrons by metals, easily recognized and
studied by students, it is not the only phenomenon in which
electrons are emitted by metals. Electrons can be emitted by metals
at room temperature by the application of an external electric field
$\mathcal{E}$. To contrast with the emission of electrons when a
metal is heated, this phenomenon is termed \textbf{cold emission}.
It was first explained by Fowler and Nordheim \cite{Fowler}.

When an external field is applied, an electron in the metal sees a
potential
 \begin{equation}
   U(x) ~=~ U_0-e\mathcal{E}\, x~,
 \label{eq:example}
 \end{equation}
where $x$ is the distance from the wall of the metal. This
description ignores the fact that a positive image charge will
appear at the surface of the metal as the electron is removed, and
thus an additional Coulomb attraction will be established. For an
electron of energy $E$ we find, using Gamow's formula
(\ref{eq:Gamow}),
 \begin{equation}
    T(E) ~=~  e^{-a\,(U_0-E)^{3/2}}~,
 \label{eq:Fowler}
 \end{equation}
 with
 $$
    a ~=~ {4\sqrt{2m}\over 3e\mathcal{E}\hbar}~.
 $$
 Equation (\ref{eq:Fowler}) is known as the \textbf{Fowler-Nordheim} equation.
 The quantity $U_0-E$ is known as the \textbf{work function}.

\begin{figure}[h!]
\begin{center}
    \psfrag{0}{$0$}
    \psfrag{E}{$E$}
    \psfrag{w}{W}
    \psfrag{x}{$x$}
    \psfrag{U0}{$U_0$}
    \psfrag{U}{$U$}
    \psfrag{U-eEx}{$ U(x) ~=~ U_0-e\mathcal{E}\, x~$}
    \includegraphics[width = .5 \textwidth]{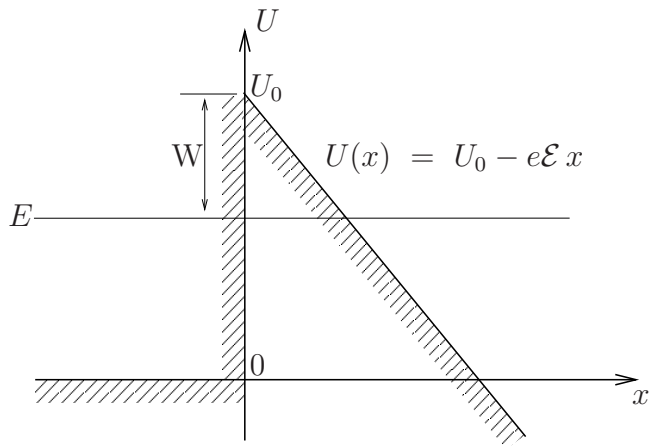}
 \caption{Model potential for cold emission.  $E$ is typically taken as the fermi energy.
 $W$ is the work function of the metal.}
 \label{fig:triangular}
\end{center}
\end{figure}

We shall now assume that the Fowler-Nordheim equation is
known---say from experimental data. Can we find the potential that
reproduces it?

According to our formula (\ref{eq:inversion})
$$
 x_1(U)-x_2(U) = - {2\over \pi e\mathcal{E}}
      \int_U^{U_0} \sqrt{U_0-E\over E-U}\,\mathrm{d}E ~.
$$
This is an elementary integral and we compute it in the appendix.
The result is
$$
 x_1(U)-x_2(U) = - {1\over e\mathcal{E}}\,(U_0-U)~.
$$
The reader may believe that we have recovered the potential
(\ref{eq:example}). Unfortunately, this is not the case as any two
functions $x_1(U)$ and $x_2(U)$ that differ by the above amount are
solutions of the inverse problem. The the cold emission potential is
recovered if we assume that $x_1(U)=0$.

%%%%%%%%%%%%%%%%
\subsection{The Issue of Uniqueness}

The reader, at this point, may believe that the apparent conflict of
our result with that which would have obtained by the method of
Gel'fand-Levitan is due to the approximations used to produce
formula (\ref{eq:Gamow}). However, a moment's reflection will reveal
that this cannot be so. The mathematical statement of the problem is
independent of the underlying physics which can be left aside.

The answer to this puzzle is quite simple.  The method of Gel'fand
and Levitan makes use of the amplitude $b(k)$ (in the situation were
considering there is no bound spectrum) which is a \textit{complex}
quantity. However, in our case we make use of $T$ which is a
\textit{real} quantity---$T=|a(k)|^2$ with $|a(k)|^2+|b(k)|^2=1$.
Thus, we have lost information about regarding phase.

Given the transmission coefficient $T(E)$, $b(k)$ can be anything of
the form
\begin{displaymath}
 b(k) ~=~ \sqrt{1-T(E(k))} ~ e^{i f(k)}~,
\end{displaymath}
where $f(k)$ is a real-valued function.
 Each  distinct potential among the family of our solutions,
corresponds to a different choice of $f(k)$.

%%%%%%%%%%%%%%%%%%%%%%%%%%%%%%%%%%%%%%%%%%%%%%%%%%%%%%%%%%%%%%%%%%%%%%%%%%
\section{Discussion and Conclusion}

We have found that, in quantum mechanics, $T(E)$, the probability
for transmission and the analog of the classical scattering data,
does not uniquely determine the potential, just as it is in
classical mechanics. However, quantum mechanics does afford us an
additional set of data, the phase difference $f(k)$, which
corresponds to measurements of time delay (p.138 of \cite{delay}).
It is only with both these sets of data that we can uniquely
determine the potential.

The approximate nature of Gamow's formula is an irrelevant feature
for the problem we have studied. However, a different kind of
question may be asked which makes this feature relevant: Assuming
that the potential barrier is even and a unique solution $U(x)$ can
be found, what is the error in determining the potential? That is,
how close is the solution $U(x)$ to the real potential,
$U_{real}(x)$, that gave the experimental data $T(E)$? This question
remains open.

Finally, toward proving our result, we have succeeded in solving a
modified version of Abel's equation: Given the integral equation,
$$
  \int_E^a {\phi(U)\over\sqrt{U-E}}\,dU ~=~ f(E)~,
$$
where $f(E)$ is a known function and $\phi(U)$ is an unknown
function, we have shown that the solution is given by
$$
  \phi(U) ~=~ -{1\over\pi} \, {d\over dU} \,
              \int_U^a {f(E)\over\sqrt{E-U}}\,dE~.
$$

%%%%%%%%%%%%%%%%%%%%%%%%%%%%%%%%%%%%%%%%%%%%%%%%%%%%%%%%%%%%%%%%%%%%%%%
\section*{Acknowledgements}
This research was sponsored by generous grants from the University
of Central Florida Honors College and Office of Undergraduate
Studies. S.G. would like to thank Rick Schell and Alvin Wang for
this financial support.  S.G. would like to thank C.E. for the
opportunity to work on this project as well as his generous time and
guidance throughout this work and S.G.'s studies.

%%%%%%%%%%%%%%%%%%%%%%%%%%%%%%%%%%%%%%%%%%%%%%%%%%%%%%%%%%%%%%%%%%%%%%%%%%
\section*{Appendix}

The integrals
 \begin{eqnarray*}
   I = \int {\mathrm{d}E\over\sqrt{(\beta-E)(E-\alpha)}}~, \quad
   J = \int {\sqrt{\beta-E}\over\sqrt{E-\alpha} }\,\mathrm{d}E
 \end{eqnarray*}
which have appeared in our article are elementary, but their
calculation is somewhat lengthy. We present their calculation here.

Introducing the substitution
$$
  u^2 ~=~ \sqrt{\beta-E\over E-\alpha}
$$
in $J$, we can rewrite it as
$$
  J ~=~ -2(\beta-\alpha)\,\int {u^2\over (1+u^2)^2}\,du
    ~=~ -2(\beta-\alpha)\,\left\lbrack
                           \int {1\over 1+u^2}\,du
                          -\int {1\over (1+u^2)^2}\,du
                           \right\rbrack~.
$$
However,
$$
     {1\over\lambda} \tan^{-1}{u\over\lambda} ~=~
     \int {1\over\lambda^2+u^2}~.
$$
Therefore,
$$
   J ~=~ -2(\beta-\alpha)\,\left\lbrack \tan^{-1}u
         -{1\over2}{\partial\over\partial\lambda}\left({1\over\lambda}
         \tan^{-1}{u\over\lambda}\right)\Big|_{\lambda=1}
                           \right\rbrack~,
$$
and, finally,
$$
  J ~=~ \sqrt{(E-\alpha)(\beta-E)}
     -(\beta-\alpha)\,\tan^{-1}\sqrt{\beta-E\over E-\alpha}~.
$$
It is immediate that
$$
  \int_\alpha^\beta \sqrt{\beta-E\over E-\alpha}\,\mathrm{d}E
  ~=~ (\beta-\alpha)\,{\pi\over2}~.
$$
Using Leibnitz's rule for the differentiation of integrals depending
on a parameter, we can also easily obtain
$$
 \int_\alpha^\beta {\mathrm{d}E\over\sqrt{(\beta-E)(E-\alpha)}}
 ~=~ 2\,{\partial I\over\partial\beta}
 ~=~ \pi~.
$$

%%%%%%%%%%%%%%%%%%%%%%%%%%%%%%%%%%%%%%%%%%%%%%%%%%%%%%%%%%%%%%%%%%%%%%%%%%%

\end{document}